\documentclass[11pt,twoside]{article}

\usepackage{asp2014}

\aspSuppressVolSlug
\resetcounters

\begin{document}

\title{Modern Python at the Large Synoptic Survey Telescope}

\author{Tim~Jenness$^1$
\affil{$^1$Large Synoptic Survey Telescope, Tucson, AZ, USA; \email{tjenness@lsst.org}}}

\paperauthor{Tim~Jenness}{tjenness@lsst.org}{0000-0001-5982-167X}{LSST}{Data Management}{Tucson}{AZ}{85719}{USA}

\begin{abstract}
  The LSST software systems make extensive use of Python, with almost all of it initially being developed solely in Python 2.
  Since LSST will be commissioned when Python 2 is end-of-lifed it is critical that we have all our code support Python 3 before commissioning begins.
  Over the past year we have made significant progress in migrating the bulk of the code from the Data Management system onto Python 3.
  This poster presents our migration methodology, and the current status of the port, with our eventual aim to be running completely on Python 3 by early 2018.
  We also discuss recent modernizations to our Python codebase.
\end{abstract}

\section{Introduction}

Python 3.0 was released in December 2008, just after the release of Python 2.6, with Python 2.7 first being released in July 2010.
The development of the Data Management software for the Large Synoptic Survey Telescope \citep[LSST;][]{2008arXiv0805.2366I} began in 2004 \citep{2004AAS...20510811A}, around the time Python 2.4 was released, and was exclusively using Python 2 until work on supporting Python 3 began in 2016 \citep{P6-12_adassxxvi}.
At LSST we use many packages that have pledged to drop support for Python 2 no later than 2020,\footnote{\url{http://www.python3statement.org}} including Astropy \citep{2013A&A...558A..33A}, pandas, matplotlib, and Jupyter.
Many of those packages are dropping Python 2 in the near term and are pledging to do minimal support of Python 2 in their Long Term Support releases.
The combination of no new features in these package from 2018 with the commissioning time line for LSST, implies that we should switch to Python 3 sooner rather than later; switching just before, or even during, commissioning starts would cause major complications.

\section{Python 3 Porting}

We used the Python \verb|future| package\footnote{\url{http://python-future.org}} to port Python 2 software to Python 3 and to aid development during the time that both Python 2 and 3 are being supported.
We chose \verb|future| because it is designed to make compatible code look like it is written for Python 3 natively and minimizes explicit checks for Python version.
The \verb|futurize| command worked really well and support for the two stage conversion was important.
Stage 1 modernizes code to use Python 2.7 constructs such as modern exception catching, checking if a key is in a \texttt{dict} using \verb|in| rather than \verb|has_key|, and use of \verb|__future__| for print function.
Stage 2 does more extensive rewrites to support Python 3 changes to builtins and the standard library.
It also replaces \verb|map(filter(lambda())| constructs with more readable list comprehensions.
We found that using the Unicode \verb|str| object provided by \verb|future| to emulate the Python 3 \verb|str| in Python 2 added more complication than was desired so we have left string objects as their native type in many places.
When supporting 2 and 3 in a single codebase, there are times when the code has to include Python 2 types such as \texttt{basestring} and \texttt{long} to allow the Python 2 code to work correctly even though in Python 3 these types are meaningless.
The \verb|future| package provides the \texttt{past.builtins} package to support this, although using it is a hint to the future that the code will have to be modified when Python 2 support is dropped.
Proper handling of bytes and strings is the biggest headache when switching Python versions, since Python 2 is much more relaxed about string encodings and code that works fine on Python 2 can cause surprising errors on Python 3.
\citet{SQR-014} provides a more detailed description of the migration process.
The LSST baseline Python 3 is version 3.6.

\section{LSST Python Software}

Python is used widely at LSST with the key software being:
Science Pipelines\footnote{For an overview of the LSST pipeline from the perspective of HSC see \citet{2017arXiv170506766B}.};
VO and Data Access Services;
simulations software \citep{2014SPIE.9150E..14C};
LSST Scheduler \citep{2016SPIE.9910E..13D}; and
wavefront sensor data processing \citep{2016SPIE.9906E..3BT}.
The biggest user of Python is in Data Management \citep[DM;][]{O3-1_adassxxv} and the bulk of this software was ported to Python 3 in the second half of 2016.
DM are regularly running Python 3 now for science pipelines, data access services and the Qserv database administration scripts.
DM currently use some DESDM infrastructure at NCSA \citep{2012SPIE.8451E..0DM}; those will be updated to work with Python 3 by early 2018.
The ``sims'' software has been ported to Python 3 and has no remaining issues.
The Scheduler software has had an initial modernization pass to support Python 3 but has not yet been tested.
The Scheduler uses the Software Abstraction Layer \citep[SAL;][]{2016SPIE.9906E..5CM} message system (based on DDS) \citep{2016SPIE.9911E..25R} and the Python 3 bindings were not available until October 2017.
We expect the Scheduler to be running with Python 3 by the end of the 2017.
The wavefront sensor processing software uses DM software and will be running Python 3 when it is completed.

\section{Data Management and Simulations Software}

In addition to supporting Python 3, we have made two major improvements to the Python infrastructure for DM and Simulations.
We have replaced our SWIG bindings \citep{beazley2003automated} with \texttt{pybind11}\footnote{\url{http://pybind11.readthedocs.io}}, and we have migrated our test execution environment to use \texttt{pytest}\footnote{\url{https://pytest.org/}}.

The \texttt{pybind11} project provides a native C++ mechanism for building Python interfaces.
Unlike SWIG, these interfaces are not constructed automatically and have to be constructed manually.
The promise of automated interface building was never really fulfilled and we previously had extensive SWIG customizations to make it work with our codebase and the interfaces were not "Pythonic", with substantial Python interface code hidden in SWIG files.
In some packages we now have less interface code and in others we have a little more, but the new interfaces have a clear definition that is understandable and maintainable.
In particular keyword arguments are now significantly easier to deal with.

Previously, for our unit tests, the build system would run each test script with the Python interpreter and look for the exit status of the script.
This worked, but the execution framework had almost no visibility into the tests themselves.
In particular there was no way to determine how many tests ran or how many were skipped, and there were no metrics reported on the execution times of individual tests.
We now use \texttt{pytest} to run the tests and write the test output into a file compatible with JUnit so that our continuous integration (CI) system can display the results.
We make use of a number of plugins including \texttt{pytest-xdist} to run the tests in parallel, and \texttt{pytest-flake8} to run the \texttt{flake8} linter on the source code.
This ability to test for style compliance is important since we have recently adopted the Python PEP8 style with some minor exceptions.
The \texttt{pytest-cov} plugin is being investigated to enable coverage reports from the unit tests.
One difference between the old and new approaches is that \texttt{pytest} runs all the tests in a shared environment either in a single process or a small number of parallel subprocesses.
This did require that some tests were modified to ensure that global state was rest at the end of each test, and some tests did have to be modified to ensure that the random number seed was in a fixed state at set up.
One complication with \texttt{xdist} is that we had to be careful with file system temporary storage since tests from the same test class can be scheduled on different processes, leading to races if fixed output files or directories are used.
Some tests had to be refactored to use temporary output files or directories to prevent races.

Version 14.0 of the DM Science Pipelines Stack was released in October 2017.\footnote{See \citet{P056_adassxxv} for an overview of version 11.0.}
This is the first release using pybind11.
This release also begins to use the Starlink AST library \citep{2016A&C....15...33B} for WCS handling, allowing complex WCS solutions to be created by combining discrete mappings in series or in parallel, in a manner that is not currently supported by the FITS WCS standard.
Full release notes can be found at: \url{https://pipelines.lsst.io/releases/notes.html}.

\section{Dropping Python 2}

We have been supporting Python 3 and Python 2 in the Data Management software since Summer 2016, and this has given our user community time to become accustomed to Python 3.
There is a cost to supporting Python 2, with extra CI resources, source code distractions with constructs that are not needed in Python 3, Python 3 features that cannot be used, and developers either running with two local installations or discovering late that their working code fails on Python 2.
To simplify our development roadmap, and give clarity to the community, LSST DM will drop support for Python 2 following the release of v15.0 of the DM  Pipelines Stack in Spring 2018.
We will support critical bug fixes to v15.0 until the release of v16.0 in late 2018. This timeline is consistent with the release of Astropy v3.0; their first release without Python 2 support \citep{APE10}.

\section{Summary}

LSST will be using Python 3.6 internally starting in 2018.
This includes data challenges, and integration and testing activities.
The Data Management software will drop support for Python 2 following the release of v15.0 in Spring 2018.

\acknowledgements I thank Nate Lust for his manuscript review. This material is based upon work supported in part by the National Science Foundation through Cooperative Agreement 1258333 managed by the Association of Universities for Research in Astronomy (AURA), and the Department of Energy under Contract No. DE-AC02-76SF00515 with the SLAC National Accelerator Laboratory. Additional LSST funding comes from private donations, grants to universities, and in-kind support from LSSTC Institutional Members.

\end{document}